\documentclass[a4paper,10pt,twoside]{cpc-hepnp}
\usepackage{CJK,upgreek,fancyhdr}
\usepackage{multicol}
\usepackage{graphicx}
\usepackage{booktabs}
\usepackage{amssymb,bm,mathrsfs,bbm,amscd}
\usepackage[tbtags]{amsmath}
\usepackage{lastpage}

\begin{document}
\begin{CJK*}{GB}{gbsn}

\fancyhead[c]{\small Chinese Physics C~~~Vol. xx, No. x (201x) xxxxxx}
\fancyfoot[C]{\small 010201-\thepage}

\footnotetext[0]{Received 31 June 2015}

\title{The impact of the off-diagonal cross-shell interaction on $^{14}$C\thanks{Supported by the National Natural Science Foundation of China (11305272), the Special Program for Applied Research on Super Computation of the NSFC Guangdong Joint Fund (the second phase), the Guangdong Natural Science Foundation (2014A030313217), the Pearl River S\&T Nova Program of Guangzhou (201506010060), the Tip-top Scientific and Technical Innovative Youth Talents of Guangdong special support program (2016TQ03N575), and the Fundamental Research Funds for the Central Universities (17lgzd34).}}

\author{%
      Cen-Xi Yuan (Ô¬á¯Ïª)$^{1;1)}$\email{yuancx@mail.sysu.edu.cn}%
}
\maketitle

\address{%
$^1$ Sino-French Institute of Nuclear Engineering and
Technology, Sun Yat-Sen University, Zhuhai 519082, China\\
}

\begin{abstract}
The shell-model investigation is performed to show the impact on the structure of $^{14}$C contributed by the off-diagonal cross-shell interaction, $\langle pp \mid$V$\mid sdsd\rangle$, which represents the mixing between the $0$ and $2\hbar\omega$ configurations in the $psd$ model space. Observed levels of the positive states in $^{14}$C can be nicely described in $0-4\hbar\omega$ or a larger model space through the well defined Hamiltonians, YSOX and WBP, with a reduction on the strength of the $\langle pp \mid$V$\mid sdsd\rangle$ interaction in the latter. The observed B(GT) values for $^{14}$C can be generally described by YSOX, while WBP and their modifications on the $\langle pp \mid$V$\mid sdsd\rangle$ interaction fail for some values. Further investigation shows the effect of such interaction on the configuration mixing and occupancy. The present work shows examples on how the off-diagonal cross-shell interaction strongly drives the nuclear structure.
\end{abstract}



\footnotetext[0]{\hspace*{-3mm}\raisebox{0.3ex}{$\scriptstyle\copyright$}2013
Chinese Physical Society and the Institute of High Energy Physics
of the Chinese Academy of Sciences and the Institute
of Modern Physics of the Chinese Academy of Sciences and IOP Publishing Ltd}%

\begin{multicols}{2}

\section{Introduction}

The investigation of the nuclear interaction is of great importance in nuclear physics. Generally speaking, two approaches are used to study the nuclear interaction, starting from the realistic nucleon-nucleon ($NN$) force and from the nuclear data including binding energies and levels. The two approaches are used together and compared with each other in various nuclear models, such as the nuclear shell model~\cite{talmi2003}.

To use the realistic $NN$ force in a shell-model investigation, two problems need to be overcome; the strong short-range repulsion and the truncated of the model space~\cite{Jensen1995,Coraggio2009}. In the construction of the effective shell-model Hamiltonian, the latter problem is normally solved through the many-body perturbation theory, which has certain difficulties in dealing with the cross-shell interaction (the interaction between different major oscillator shells)~\cite{Jensen1995}. Recently, the extended Kuo-Krenciglowa (EKK) method is suggested to derive the effective interaction among several oscillator shells~\cite{Takayanagi20111,Takayanagi20112}. Its applications in the $sdpf$ region show a nice agreement with the observed data, focusing on the binding energies, the levels of two-nucleon pairs, and the energies of the first $2^{+}$ and $4^{+}$ states in the even-even nuclei~\cite{Tsunoda2014,Tsunoda2016}.

Many effective Hamiltonians are well defined considering the observed binding energies and levels, while some of them have realistic basis. Some examples are CK for $p$ shell~\cite{ck}, USD family for $sd$ shell~\cite{usd,usdab}, GXPF1 for $pf$ shell~\cite{Honma2002}, MK~\cite{mk1975}, WBT~\cite{wbt1992}, WBP~\cite{wbt1992}, SFO~\cite{suzuki2003} for $psd$ shells, and SDPF-M~\cite{Utsuno1999} for $sdpf$ shells.

There are two types of the cross-shell interaction in the two-body part of an effective Hamiltonian constructed for two major oscillator shells, which are $\langle N,N+1 \mid$V$\mid N,N+1\rangle$ and $\langle N,N \mid$V$\mid N+1,N+1\rangle$, where $N$ and $N+1$ are one and its next major oscillator shells, respectively. The first type, especially its diagonal part, is very important for the investigation of the neutron-rich nuclei, of which the protons are in $N$ shell and the neutrons are in $N+1$ ($N+2$ for some extreme cases) shell. Its strength can be determined by considering the observed data of those nuclei. However, much less knowledge is known for the second type, which is purely off-diagonal, corresponding to the mixing between $n\hbar\omega$ and $(n+2)\hbar\omega$ configurations, where $n$ means the number of nucleons excited to next major shell. Some early effective Hamiltonians were constructed without consideration of the mixing between the $0$ and $2\hbar\omega$ configurations, such as MK, WBP, and WBT. The $\langle psd \mid$V$\mid psd\rangle$ interaction in WBP is obtained from a potential fitted to the observed data~\cite{wbt1992}. The $\langle pp \mid$V$\mid sdsd\rangle$ interactions of WBP and WBT are calculated from the same potential without considering its effect on the nuclear structure~\cite{wbt1992}. Later suggested effective Hamiltonians SFO, constructed in the $0-3\hbar\omega$ model space, did not focus on the properties of this interaction, but on the spin properties of the $p$ shell nuclei~\cite{suzuki2003}.

It is seen that the off-diagonal cross-shell interaction, connecting the $n\hbar\omega$ and $(n+2)\hbar\omega$ configurations, is not well investigated in both realistic and phenomenological investigations. One reason is that its effect on nuclear structure is ``hidden'', which means it is hard to be shown in an easily understandable scheme, such as the effect of the monopole interaction on the binding energies~\cite{Duflo1995} and the shell structures~\cite{otsuka2001,otsuka2005,otsuka2010}. Recently, such multipole correlations between the normal and intruder configurations are investigated in the neutron-rich nuclei around $N=20$ and $28$ in the $sdpf$ model space~\cite{Poves2012}. The importance of the multi-$\hbar\omega$ configuration mixing has been investigated through the Sp(3, $\mathbb{R}$) shell model \cite{Launey2015}. Clear symplectic symmetry in low-lying states of $^{12}$C and $^{16}$O is shown that their NCSM wave functions can be typically projected to a few of the most deformed symplectic basis states at the level of $85\%-90\%$ \cite{Dytrych2007}. The \emph{ab initio} symmetry-adapted NCSM (SA-NCSM) results show that the multi-$\hbar\omega$ configuration mixing are important for the description of the collective modes in light nuclei, such as $^{6}$Li, $^{6}$He, and $^{8}$Be \cite{Dytrych2013,Draayer2012}. The no-core symplectic shell model (NCSpM) are used to investigate the multi-$\hbar\omega$ configuration mixing in $\alpha$-clustering substructures in the low-lying states of $^{12}$C \cite{Dreyfuss2013} and in ground state rotational bands of $^{20,22,24}$Ne, $^{20}$O, $^{20,22}$Mg and $^{24}$Si \cite{Tobin2014}.

The effective Hamiltonian YSOX for $psd$ region includes the effect of the off-diagonal cross-shell interaction based on the binding energies of B, C, N, and O isotopes from the stability line to the neutron drip line~\cite{yuan2012}. The results show that the strength of the central part of the $\langle pp \mid$V$\mid sdsd\rangle$ interaction is weaker than that of the $\langle psd \mid$V$\mid psd\rangle$ interaction. The effect of the strength of the $\langle pp \mid$V$\mid sdsd\rangle$ interaction on the low-lying levels of $^{10}$B and $^{17}$C are also presented. But the absolute values of the levels of these two nuclei do not vary significantly with the strength of the $\langle pp \mid$V$\mid sdsd\rangle$ interaction~\cite{yuan2012}. It is necessary to find more solid evidences on the effect of such interaction on the nuclear structure. For example, levels and transition rates of some states, which are dramatically influenced by the $\langle pp \mid$V$\mid sdsd\rangle$ interaction, can be well described through YSOX, and other Hamiltonians with modifications on the strength of this interaction.

A good candidate for the above considerations is $^{14}$C, one of the most widely known isotopes. The long lifetime of $^{14}$C is a long standing problem for theoretical models, which can be understood by the cancellation of the transition matrix elements between two main components in the $p$ shell~\cite{talmi2003}. A few microscopic approaches based on $NN$ (and $NNN$) force are performed to investigate the origin of the extreme small B(GT) value between the ground states of $^{14}$N and $^{14}$C~\cite{Aroua2003,Holt2008,Holt2009,Maris2011,Ekstrom2014}. It should be noted that the energies of the first few positive states of $^{14}$C and the B(GT) values for these states can not be well described in the above microscopic investigations, the antisymmetrized molecular dynamics (AMD) method~\cite{Enyo2014}, or the shell model in the $0-2\hbar\omega$ model space with existing Hamiltonians, such as WBP, SFO, and YSOX~\cite{yuan2012}. The level of $^{14}$C are normally excluded in the construction of a Hamiltonian for global $psd$ region.

The $sd$ shell configurations are obviously important for the level of $^{14}$C. The valence protons and neutrons in $^{14}$C fully occupy $Z=6$ sub shell and $N=8$ major shell, respectively. The single particle states of $^{13}$C~\cite{nndc} indicate that the excitation energies inside $p$ shell are with the same magnitude of the two-nucleon excitation energies from $p$ to $sd$ shell. Our recent conference proceeding~\cite{yuanNPR} shows that the first few positive states are some of the mixing between $0\hbar\omega$ ($\sim80\%$) and $2\hbar\omega$ ($\sim20\%$) configurations, $0^{+}_{1}$, $2^{+}_{1}$, and $1^{+}_{1}$ states, some of the almost pure $2\hbar\omega$ configuration, $0^{+}_{2}$ and $2^{+}_{2}$ sates, and the pure $2\hbar\omega$ configuration, $4^{+}_{1}$ state, which is not possible to be coupled inside $p$ shell. The excitation energies of $0^{+}_{2}$, $0^{+}_{3}$, and $4^{+}_{1}$ states can be well described in a simple model of $(sd)^{2}$ states~\cite{Fortune2011,Fortune2014}.

In this paper, the structure of $^{14}$C is investigated in the framework of the shell model up to the $6\hbar\omega$ excitation. It is shown that both the strength of the $\langle pp \mid$V$\mid sdsd\rangle$ interaction and the inclusion of the $4\hbar\omega$ configuration are crucial to reproduce the observed data of $^{14}$C. The details of the Hamiltonian used in the present work is briefly introduced in Section 2. Levels and transition rates are discussed in Section 3 and 4, respectively.  Some further discussions are presented in Section 5.

\section{Hamiltonian}

The nuclear shell model is widely used to investigate the structure of light and medium mass nuclei~\cite{brown2001,Caurier2005}. In the $psd$ region, the Hamiltonians MK, WBT, and WBP are the successful ones. Both the $\langle pp \mid$V$\mid sdsd\rangle$ and $\langle pp \mid$V$\mid sdsd\rangle$ interactions in WBP are calculated through the same potential, which is convenient for discussing the different strengths in the present study. Thus WBP and its modifications on $\langle pp \mid$V$\mid sdsd\rangle$ interaction are considered in the following discussions.

The recent suggested Hamiltonian YSOX is also used in the present investigation. The $\langle pp \mid$V$\mid pp\rangle$ and $\langle sdsd \mid$V$\mid sdsd\rangle$ parts of YSOX are from the corresponding parts of SFO and SDPF-M, respectively~\cite{yuan2012}. The two types of the cross-shell interaction, $\langle psd \mid$V$\mid psd\rangle$ and $\langle pp \mid$V$\mid sdsd\rangle$, are calculated through V$_{MU}$~\cite{otsuka2010} plus spin-orbit force from M3Y~\cite{m3y1977}. V$_{MU}$ is the monopole based universal interaction including a Gaussian type central force and a $\pi+\rho$ bare tensor force, which assumes that the renormalization effect is mostly included in the central force~\cite{otsuka2010,tsunoda2011}. The validity of taking V$_{MU}$ plus spin-orbit term as the cross-shell interaction in the shell model is examined in various works besides $psd$ region, such as $sdpf$ region~\cite{utsuno2012,Utsuno2015} and $pfsdg$ region~\cite{Togashi2015}. Such nuclear force is used to estimate the reduction effect caused by the weakly bound proton $1s_{1/2}$ orbit~\cite{yuan2014} and taken as the cross-shell interaction between two major shells, $Z=28$-$50$ and $N=82$-$126$ shells~\cite{yuan20162}. The first $19/2^{-}$ state in $^{129}$Pd is predicted to be a possible neutron-decaying isomer~\cite{yuan20162}.

Among the above applications of V$_{MU}$, the renormalization effect is assumed to be contributed mostly by the central part. The tensor and spin-orbit parts are calculated through unchanged strength for both $\langle psd \mid$V$\mid psd\rangle$ and $\langle pp \mid$V$\mid sdsd\rangle$ interactions. The strengths of the central part of $\langle psd \mid$V$\mid psd\rangle$ and $\langle pp \mid$V$\mid sdsd\rangle$ interactions in YSOX are $0.85$ and $0.55$ of the original one, respectively~\cite{yuan2012}. The much weaker strength of $\langle pp \mid$V$\mid sdsd\rangle$ interaction gives nice descriptions on the binding energies of B, C, N, and O isotopes. The effect of the strength of a such central force is shown for low lying levels of $^{10}$B and $^{17}$C, but the change is not remarkable ($\leq 0.5$ MeV), when the strength varies from $0.55$ to $0.85$ (YSOX$+$) or $0.25$ (YSOX$-$) of the original value~\cite{yuan2012}. The Hamiltonian YSOX$+$, with the same strength for two types of the cross-shell interaction is also used in the present work for comparison.

The two body matrix elements (TBME) of WBP are compared in Ref.~\cite{yuan2012} with those of YSOX for central, spin-orbit, tensor and total interactions through the spin-tensor decomposition method~\cite{kirson1973}. It is found that the two types of cross-shell interaction are quite different between YSOX and WBP, except the spin-orbit one. A modified version WBP$-$ is introduced by multiplying a factor $0.6$ to all central TBME of the $\langle pp \mid$V$\mid sdsd\rangle$ interaction in WBP, which is similar to the reduction in YSOX.

The Hamiltonians, WBP, WBP$-$, YSOX$+$, and YSOX, are used in the following discussions for comparisons with each other. The shell-model calculations are performed with newly developed code KSHELL~\cite{Shimizu2013}. The center-of-mass (c.m.) correction is needed for multi-shell calculations. The standard method suggested by Gloeckner and Lawson~\cite{cm1974} is used for the c.m. correction, which defined $H'=H_{SM}+\beta H_{c.m.}$, where $H_{SM}$ and $H_{c.m.}$ are shell-model and c.m. Hamiltonians, respectively. In the present study, $\beta=10$ is adopted.

\section{level of $^{14}$C}

\end{multicols}
\ruleup
\begin{center}
\includegraphics[scale=0.5]{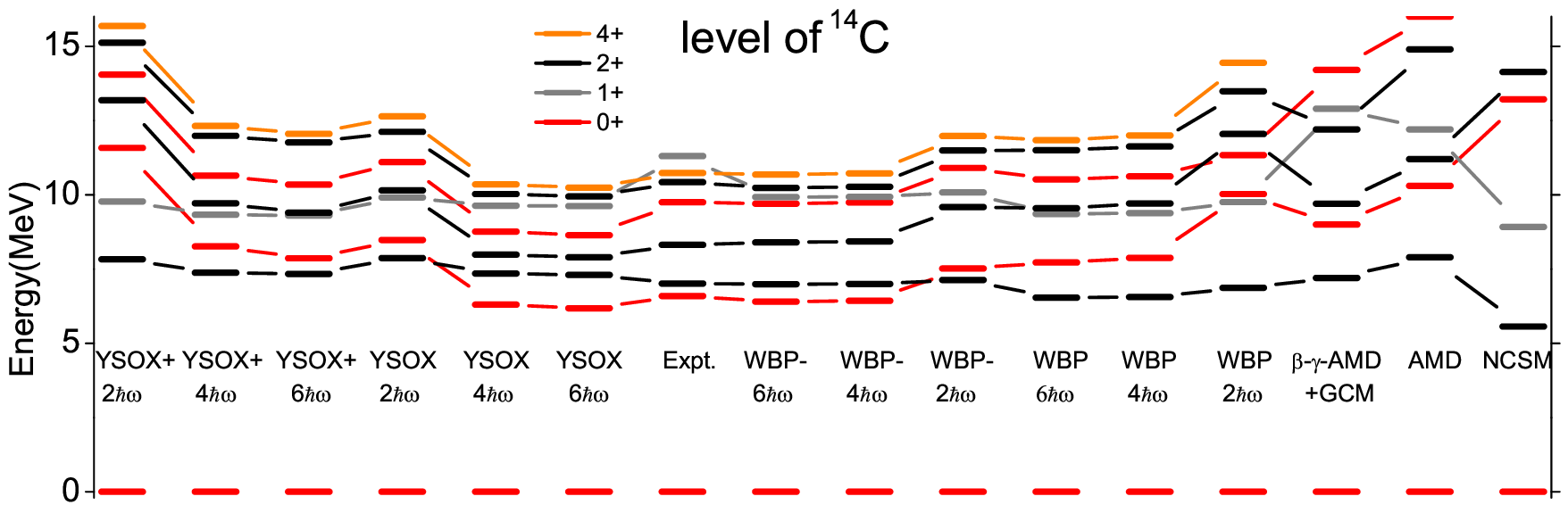}
\figcaption{\label{level1} (Color online) Comparison of the energies of the positive states in $^{14}$C between the observed data and the results of various models. Observed data, AMD, and NCSM results are taken from Ref.~\cite{nndc},~\cite{Enyo2014}, and \cite{Aroua2003}, respectively}
\end{center}
\ruledown
\begin{multicols}{2}

The first few positive states of $^{14}$C are not well described through various models because of the large shell gaps for both protons and neutrons. Some results obtained from NCSM~\cite{Aroua2003} and AMD~\cite{Enyo2014} are compared with those from the shell model in Fig.~\ref{level1}. None of the previous works and the shell-model results up to the $2\hbar\omega$ model space can reproduce the correct order of the first few positive states. For example, $2^{+}$ state is calculated to be the first excited positive state instead of the observed $0^{+}$ state. It should be noted that both YSOX and WBP can give nice description on the levels, moments, and transition rates of the nearby nuclei. When the model space is up to $4\hbar\omega$ in the shell-model calculations, significant changes in the levels are found compared with those from the $0$-$2\hbar\omega$ model space. The excitation energies of $0^{+}_{2,3}$, $2^{+}_{2,3}$, and $4^{+}_{1}$ are dramatically dropped down due to the inclusion of the $4\hbar\omega$ model space, while those of the $2^{+}_{1}$ and $1^{+}_{1}$ states vary little. The first five and the last two states are dominated by the $2\hbar\omega$ and $0\hbar\omega$ configurations, respectively. Such results agree with the analysis on $0^{+}_{2,3}$ and $4^{+}_{1}$ states based on $(sd)^{2}$ configuration~\cite{Fortune2011,Fortune2014}, $0^{+}_{2,3}$ and $2^{+}_{2,3}$ states based on AMD method~\cite{Enyo2014}. It should be mentioned that the $2^{+}_{1}$ state obtained from YSOX in $0$-$4\hbar\omega$ model space has very strong mixing between the $0\hbar\omega$ and $2\hbar\omega$ configurations, which will be discussed in Section 5.

It is seen that the inclusion of the $4\hbar\omega$ model space is not enough to obtain a nice description on the energies of positive states in $^{14}$C through the Hamiltonians, YSOX$+$ and well defined WBP. Both of them have the same strength in two types of the cross-shell interaction, $\langle psd \mid$V$\mid psd\rangle$ and $\langle pp \mid$V$\mid sdsd\rangle$. The Hamiltonians, YSOX and WBP$-$, can well describe these energies with just one modification, weakening of the $\langle pp \mid$V$\mid sdsd\rangle$ interaction. All shell-model results for the energy of the $1^{+}_{1}$ state are lower than the observed value. This state is dominated by the excitation inside $p$ shell, rarely influenced by the higher $\hbar\omega$ excitation and the $\langle pp \mid$V$\mid sdsd\rangle$ interaction. Its excitation energy is not further discussed in the present work.

\begin{center}
\includegraphics[scale=0.25]{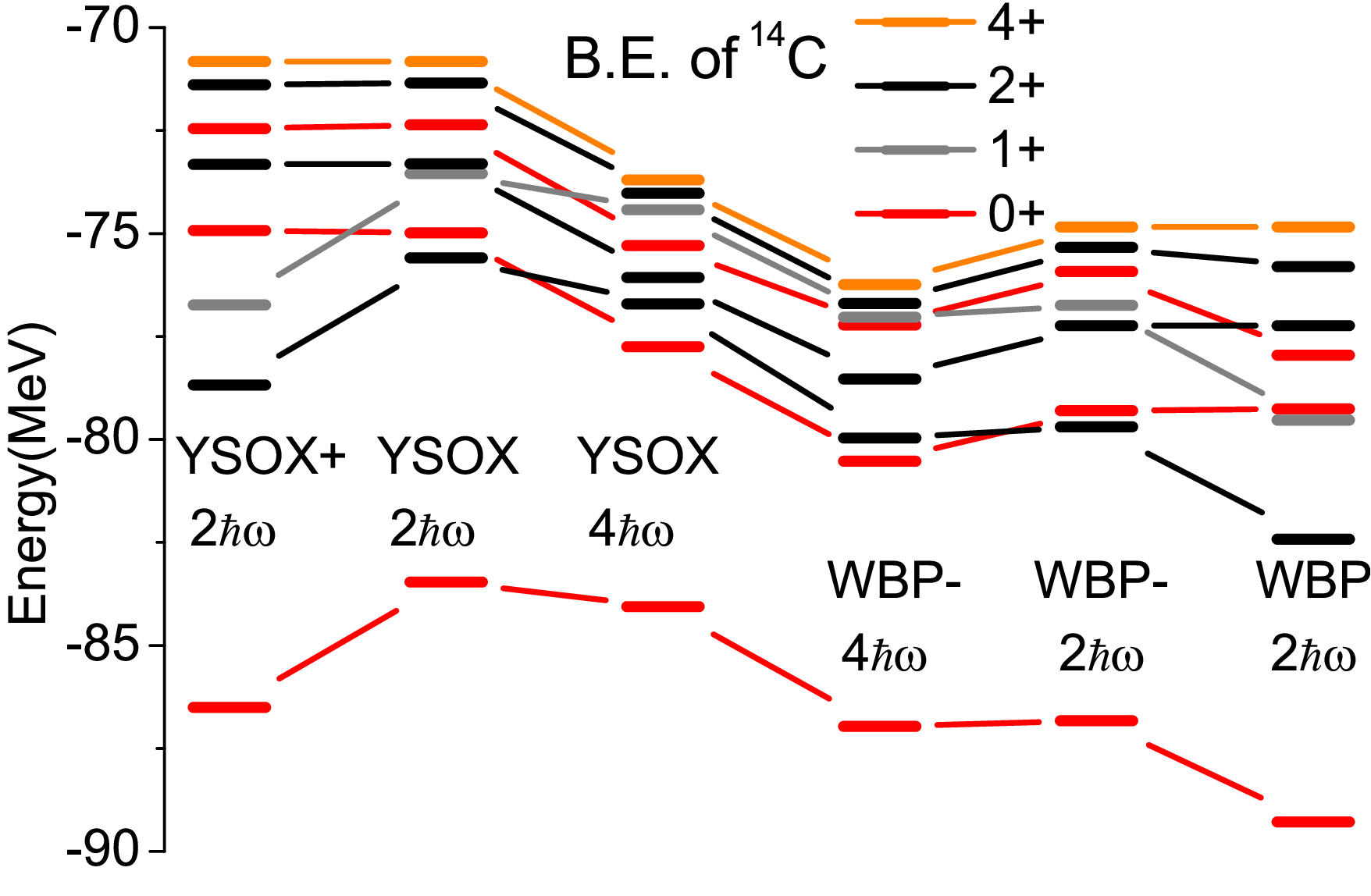}
\figcaption{\label{level2} (Color online) Comparison of the binding energies of the positive states in $^{14}$C among various shell-model calculations.}
\end{center}

Figure~\ref{level2} presents the binding energies of each states, which gives a clear view on how the $\langle pp \mid$V$\mid sdsd\rangle$ interaction and the $4\hbar\omega$ configuration drive the evolution of the energies. It is seen that the binding energies of $0\hbar\omega$ dominated states, $0^{+}_{1}$, $2^{+}_{1}$, and $1^{+}_{1}$, rise due to the weakening of the $\langle pp \mid$V$\mid sdsd\rangle$ interaction, while other states keep almost unchanged. The reason is that the $0\hbar\omega$ dominated states include certain percentages of the $2\hbar\omega$ configuration, while the $2\hbar\omega$ dominated states include rather few percentages of the $0\hbar\omega$ configuration. The energies contributed by the mixing between $0$ and $2\hbar\omega$ configurations are weaker when the strength of the $\langle pp \mid$V$\mid sdsd\rangle$ interaction is reduced.

The inclusion of the $4\hbar\omega$ configuration leads to a different effect. The wave function of $^{14}$C up to the $4\hbar\omega$ excitation is simply written as, $a|(sd)^{0}\rangle+b|(sd)^{2}\rangle+c|(sd)^{4}\rangle$, with the three terms corresponding to the $0$, $2$, and $4\hbar\omega$ configurations, respectively. The cross-shell interaction, $\langle pp \mid$V$\mid sdsd\rangle$, connects the first two and the last two configurations, but not the first and last one, because of its two-body nature. Thus, the inclusion of the $4\hbar\omega$ configuration does not show significant effect on the $0\hbar\omega$ dominated states, but strongly affects the $2\hbar\omega$ dominated states. The further inclusion of the $6\hbar\omega$ configuration results in rather little changes in level, because these states are not dominated by the $4\hbar\omega$ configuration.

%\end{multicols}
%\ruleup
\begin{center}
\includegraphics[scale=0.3]{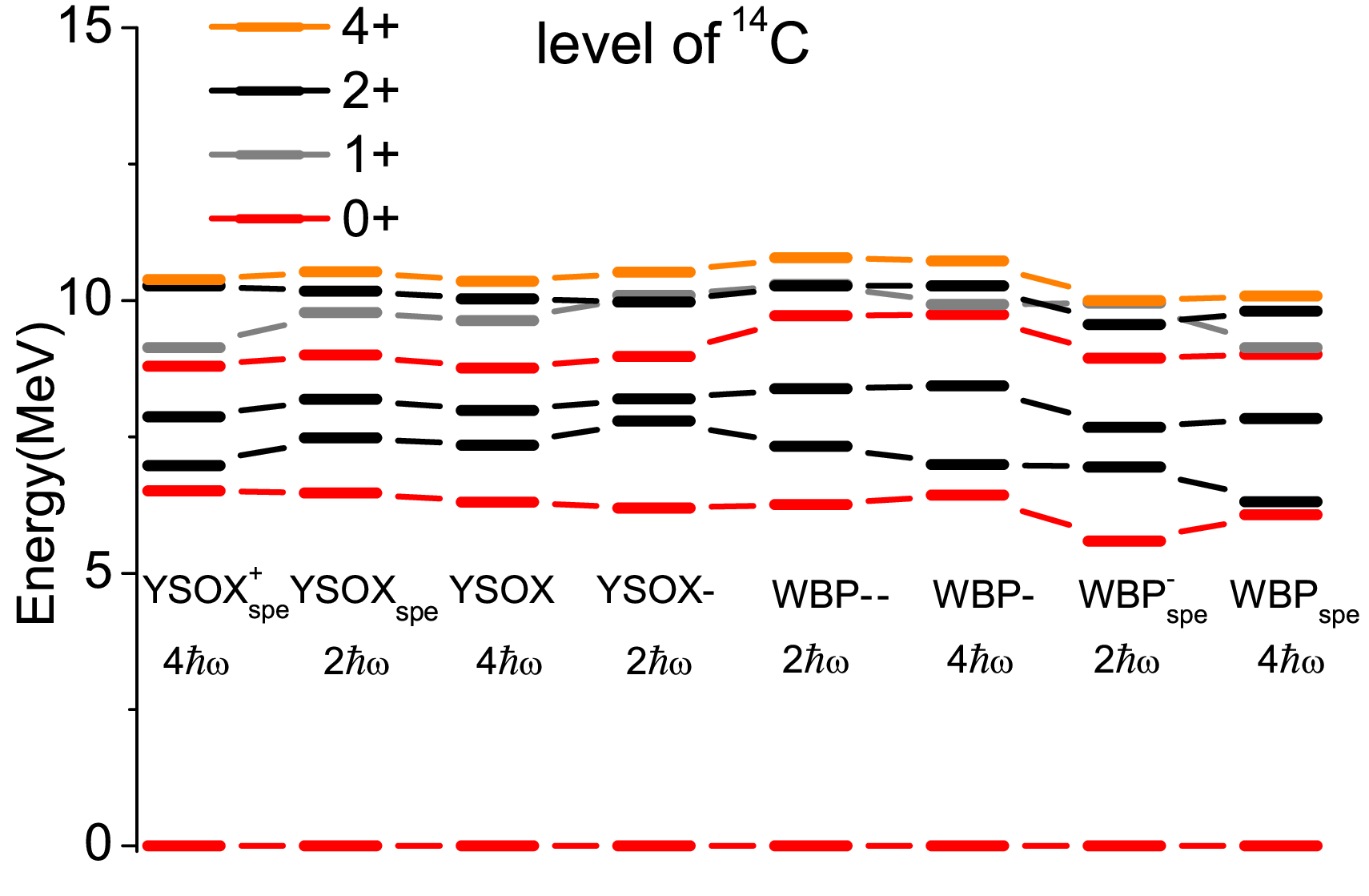}
\figcaption{\label{level3} (Color online) Comparison of the energies of the positive states in $^{14}$C among various shell-model calculations.}
\end{center}
%\ruledown
%\begin{multicols}{2}

The diagonal terms of the interaction, especially single particle energies, surely affect the level of $^{14}$C. Figure~\ref{level3} shows the effect of the single particle energies on the level of $^{14}$C. The Hamiltonians YSOX$^{+}_{spe}$ and YSOX$_{spe}$ are modified versions of YSOX+ and YSOX by reducing $1.0$ MeV on the gap between $p$ and $sd$ shell. The results indicate that the effect of reducing such gap is similar to the reduction of the $\langle pp \mid$V$\mid sdsd\rangle$ interaction and the increment of the model space from $2\hbar\omega$ to $4\hbar\omega$, by comparing among the levels from YSOX$^{+}_{spe}$ in $4\hbar\omega$, YSOX$_{spe}$ in $2\hbar\omega$, and YSOX in $4\hbar\omega$. For example, seen from the YSOX$^{+}_{spe}$ and YSOX results, the reduced gap shows similar effect to the reduction of the $\langle pp \mid$V$\mid sdsd\rangle$ interaction, which indicates the latter reduction actually attracts the $0$ and $2\hbar\omega$ configurations and increases the mixing between them. More details will be discussed in Section 5. The WBP results also show similar effect by reducing $0.9$ MeV on the same gap. Although the effect from the reduction on the gap is presented here, the reduction is not expected to be needed for YSOX in a real case. Because the gap and the strength of the cross-shell interaction are simultaneously fixed to the single particle levels of $^{17}$O, $^{15}$C, $^{13}$C, and other nuclei in the construction of the Hamiltonian YSOX. If the gap is changed, the strength of the cross-shell interaction also needs to be changed which may give worse description on global properties in nearby nuclei.

In a phenomenological view, the effect of the $\langle pp \mid$V$\mid sdsd\rangle$ interaction and the inclusion of the $4\hbar\omega$ configuration can be partially replaced by each other when concentrating on the level of $^{14}$C. Figure~\ref{level3} shows similar results between the $4\hbar\omega$ calculations and the results with a further reduction in the central force in the $\langle pp \mid$V$\mid sdsd\rangle$ interaction. The Hamiltonians YSOX$-$ and WBP$--$ mean the strengths of such central force are reduced to $0.25$ and $0.3$ of their original values in V$_{MU}$ and WBP, respectively. The results of YSOX and WBP$-$ up to the $2\hbar\omega$ model space are also similar to those of YSOX$+$ and WBP up to the $4\hbar\omega$ model space in Fig.~\ref{level1}. But such situations are not found from the YSOX and WBP$-$ up to the $4\hbar\omega$ model space to YSOX$+$ and WBP up to the $6\hbar\omega$ model space. It means that the reduction of the $\langle pp \mid$V$\mid sdsd\rangle$ interaction can not be fully replaced by the increment of the $n\hbar\omega$ excitation. It should be also noted that the strength of such central force in YSOX is considered through the binding energies for all B, C, N, and O isotopes. A much larger or smaller value of the strength is not suitable for these binding energies~\cite{yuan2012}.

In general, the energies of the first few positive states in $^{14}$C can be well reproduced in the $0-4\hbar\omega$ model space through the well defined Hamiltonians, YSOX and WBP, with a modification on the latter. Such modification do not change the nice description of WBP in its original model space, exclusion of the mixing between the $0$ and $2\hbar\omega$ states. Although some modifications shown in Fig.~\ref{level3} similarly describe the level of $^{14}$C, the importance of the $\langle pp \mid$V$\mid sdsd\rangle$ interaction seems not be substituted by other effects considering the global description on the nearby nuclei. The inclusion of the $4\hbar\omega$ in the present investigation is reasonable because of the existence of the states dominated by $2\hbar\omega$ configuration.

\section{Transition rates}

\end{multicols}
\ruleup
\begin{center}
\includegraphics[scale=0.5]{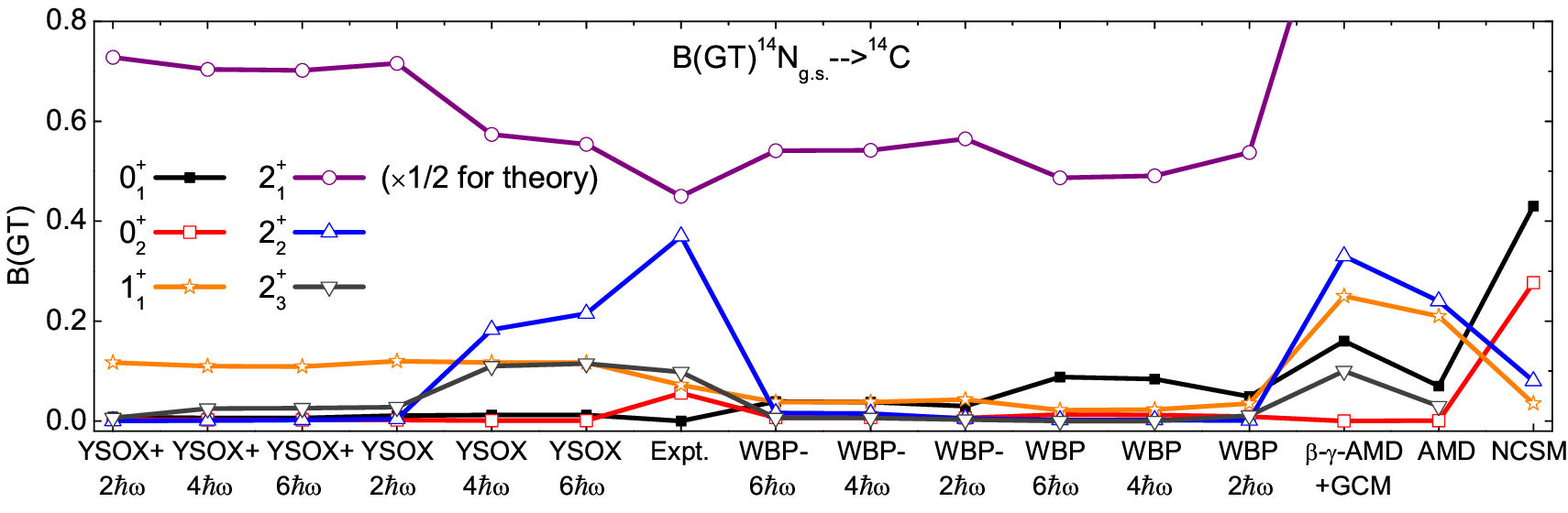}
\figcaption{\label{BGT1} (Color online) Comparison of the B(GT)($^{14}$N$_{g.s.}$ $\rightarrow$ $^{14}$C$_{0^{+},1^{+},2^{+}}$) between the observed data and the results of various models. Observed data, AMD, and NCSM results are taken from Ref.~\cite{Negret2006},~\cite{Enyo2014}, and \cite{Aroua2003}, respectively.}
\end{center}
\ruledown
\begin{multicols}{2}

Besides the energies, the transition rates also show the effect of the $\langle pp \mid$V$\mid sdsd\rangle$ interaction. Figure~\ref{BGT1} presents the B(GT) transition rates from the ground state of $^{14}$N to the $0^{+}_{1,2}$, $2^{+}_{1,2,3}$, and $1^{+}_{1}$ states of $^{14}$C. Two quenching factors $0.72$ and $0.64$ are obtained to reproduce the observed B(GT) values among the nuclei around $^{14}$C for YSOX and WBP, respectively~\cite{yuan2012}. The B(GT) values from the shell-model calculations in Fig.~\ref{BGT1} are presented with these quenching factors, $0.72$ for YSOX and YSOX$+$, $0.64$ for WBP$-$ and WBP, respectively.

The observed B(GT) values for the $0^{+}_{1}$ and $2^{+}_{1}$ states of $^{14}$C are overestimated by all theoretical results, as shown in Fig.~\ref{BGT1}. The results of the former one are not very clearly presented in the figure because of its rather small absolute value. Such small B(GT) value corresponds to the long lifetime of $^{14}$C. It is shown that the value is very sensitive to the model space and the strength of the spin-orbit and tensor force~\cite{Fayache1999}. Several of the microscopic NCSM investigations are performed to investigate the origin of the small value~\cite{Aroua2003,Holt2008,Holt2009,Maris2011}. NCSM with chiral $NN$ + $NNN$ interactions can explain the rather small transition rate \cite{Barrett2013}. In general, YSOX and YSOX$+$ give smaller B(GT) values for $0^{+}_{1}$ state compared with WBP$-$ and WBP.

The B(GT) values for $2^{+}_{1}$ state from shell model are systematically slightly larger than two times of the observed value. Certain deficiencies may exist in the descriptions of the $2^{+}_{1}$ state of $^{14}$C and/or the $1^{+}_{1}$ state of $^{14}$N. The NCSM~\cite{Aroua2003} and AMD~\cite{Enyo2014} B(GT) values for $2^{+}_{1}$ state are around five times of the observed data, which are beyond the range of Fig.~\ref{BGT1}. The B(GT) value for $1^{+}_{1}$ state is generally well described by the shell model and NCSM.

The B(GT) values for $2\hbar\omega$ dominated $2^{+}_{2,3}$ states can not be reproduced except by the AMD method and the Hamiltonian YSOX up to $4$ and $6\hbar\omega$ model spaces. The $0^{+}_{2,3}$ and $2^{+}_{2,3}$ states include cluster correlations resulting from the mixing of higher shell components in AMD calculations which can be described within the $6\hbar\omega$ model space~\cite{Enyo2014}. The present shell-model study agrees with such statement, because few differences are found on level and B(GT) values of these states between the $4$ and $6\hbar\omega$ results. But it is clearly seen that the inclusion of the higher shell components is not enough for shell model to reproduce the B(GT) values for $2^{+}_{2,3}$ states. Only the Hamiltonian YSOX can give nice descriptions on these two B(GT) values, while all others fail, such as YSOX$+$, WBP, and WBP$-$. The complicated correlation can be described by the combination of the higher excitation to $sd$ shell and the weakening of the $\langle pp \mid$V$\mid sdsd\rangle$ interaction based on V$_{MU}$ plus spin-orbit force. It is not easy to fully understand such effect, because it is difficult to know how off-diagonal interaction drives the structure of the nuclei. Some further discussions are scheduled in the next section.

\end{multicols}
\ruleup
\begin{center}
\includegraphics[scale=0.5]{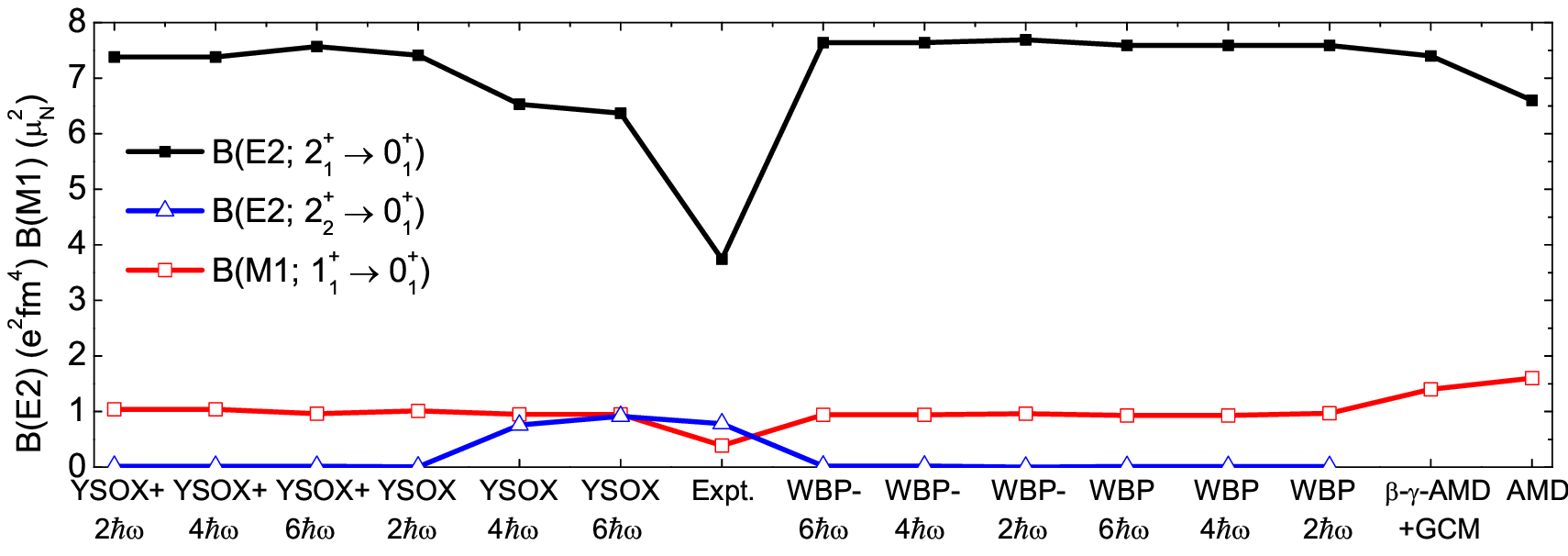}
\figcaption{\label{E2M1} (Color online) Comparison of the B(E2)($2^{+}_{1,2} \rightarrow 0^{+}_{1}$) and B(M1)($1^{+}_{1} \rightarrow 0^{+}_{1}$) in $^{14}$C between the observed data and the results of various models. Observed data and AMD results are taken from Ref.~\cite{raman2001,Selove1991} and \cite{Enyo2014}, respectively.}
\end{center}
\ruledown
\begin{multicols}{2}

Similar to the B(GT) value for $2^{+}_{1}$ state, the B(E2) value for the same state and the B(M1) value for $1^{+}_{1}$ state are also overestimated by all theoretical approaches, as shown in Fig.~\ref{E2M1}. The effective charges $e_{p}=1.27$, $e_{n}=0.23$, the effective g factors $\delta g^{(l)}_{\pi,\nu}=\pm0.1~\mu_{N}$ and $g^{(eff)}_{s}/g_{s}=0.95$ are used in the present calculations, which are obtained through the systematic trends of the electromagnetic properties of B, C, N, and O isotopes in Ref.~\cite{yuan2012}. Very few discussions are found on the B(E2) value for $2^{+}_{2}$ state, which is reported in Ref.~\cite{hayes1988}. This value is well reproduced by the results of YSOX in both the $4\hbar\omega$ and $6\hbar\omega$ model spaces, while all other calculations fail. In general, the YSOX results in $4\hbar\omega$ and $6\hbar\omega$ model spaces give better descriptions than other calculations on these transitions.

\section{Further discussions}

It is of great importance to know why YSOX gives better description than WBP$-$ on the B(GT) and B(E2) values, while both of them can reproduce the level of $^{14}$C. It should be mentioned that YSOX shows better performance than WBP in a global comparison on the B(GT) values among the nuclei around $^{14}$C~\cite{yuan2012}. The TBME and the monopole terms of YSOX and WBP are compared in Ref.~\cite{yuan2012}. Similar comparison of the $\langle pp \mid$V$\mid sdsd\rangle$ part between YSOX and WBP$-$ are presented in Fig.~\ref{ppsdsd}. Please note that the only difference between WBP and WBP$-$ is the central part of the $\langle pp \mid$V$\mid sdsd\rangle$ interaction. The comparison of the central TBME of the $\langle pp \mid$V$\mid sdsd\rangle$ interaction between YSOX and WBP$-$ shows more similarities than that between YSOX and WBP, which indicates the reason why the reduction of such central force in WBP$-$ can reproduce the level of $^{14}$C.

\begin{center}
\includegraphics[scale=0.25]{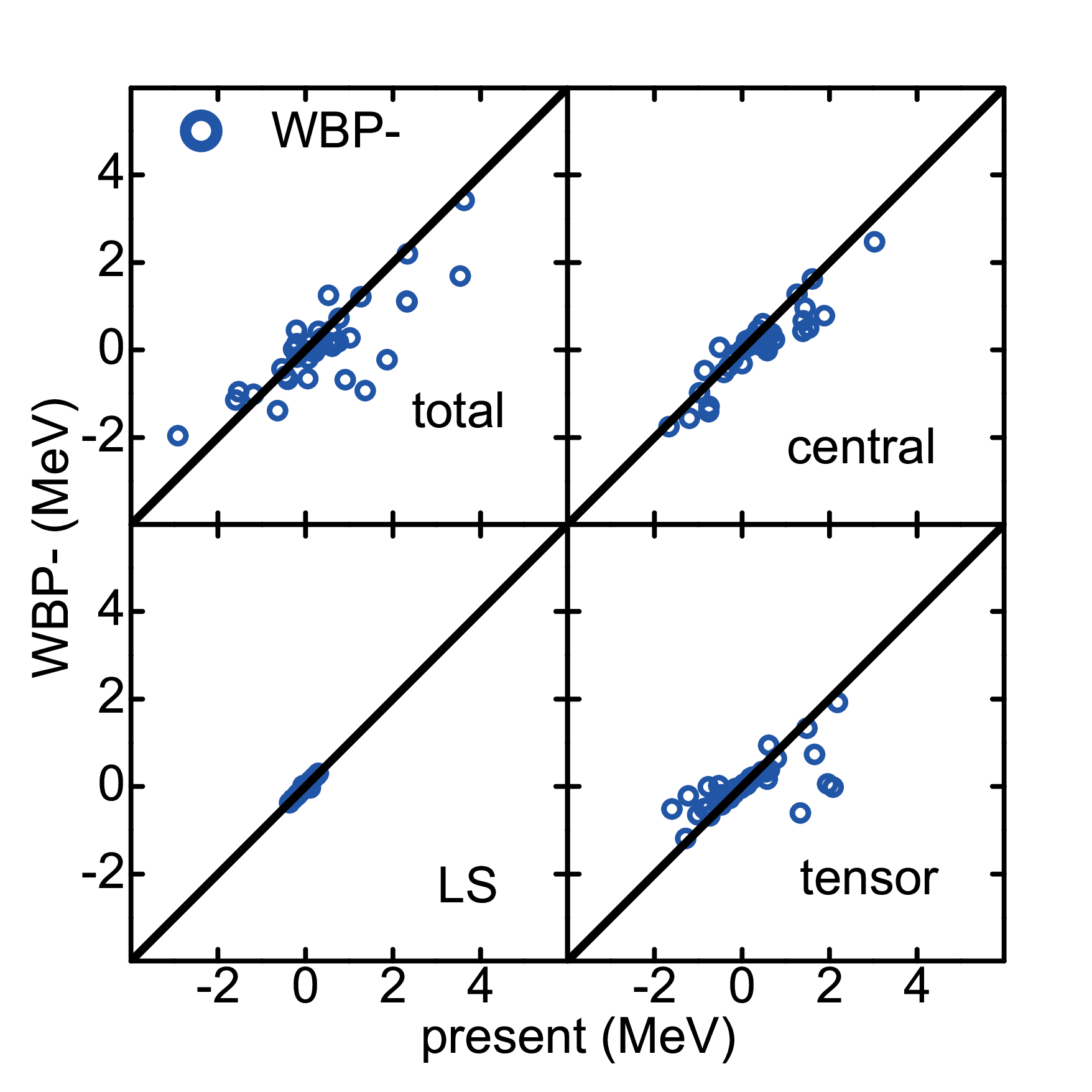}
\figcaption{\label{ppsdsd} (Color online) Comparison of the TBME of $\langle pp \mid$V$\mid sdsd\rangle$ interaction between YSOX and WBP$-$.}
\end{center}

YSOX and WBP$-$ are similar in all $\langle sdsd \mid$V$\mid sdsd\rangle$ part, central and spin-orbit forces of $\langle pp \mid$V$\mid pp\rangle$ part, spin-orbit force of $\langle psd \mid$V$\mid psd\rangle$ and $\langle pp \mid$V$\mid sdsd\rangle$ parts, seen from Ref.~\cite{yuan2012}. Most differences come from the central force of the last two parts and the tensor force of the last three parts, in total five components. The TBME of each of the five components in WBP$-$ is replaced by the corresponding one in YSOX. But none of the modified Hamiltonian can reproduce the B(GT) values for $2^{+}_{2,3}$ states. It shows the complexity of the reason that such B(GT) values can be well described by YSOX, which may be contributed by the combination of several of the five components.

\begin{center}
\tabcaption{\label{BGTdetail}The transition matrix elements of the calculated B(GT)($^{14}$N$_{g.s.}$ $\rightarrow$ $^{14}$C$_{2^{+}_{1,2,3}}$)}
\footnotesize
\begin{tabular*}{85mm}{@{\extracolsep{\fill}}ccccccccc}
\toprule  state & Hamiltonian & model space & M$_{0p_{1/2} \rightarrow 0p_{3/2}}$ &M$_{other}$ \\
\hline
$2^{+}_{1}$ &  YSOX & $2\hbar\omega$ &  1.95 & 0.13 \\
$2^{+}_{1}$ &  YSOX & $4\hbar\omega$ &  1.59 & 0.27 \\
$2^{+}_{1}$ &  YSOX+& $4\hbar\omega$ &  1.80 & 0.26 \\
$2^{+}_{1}$ &  WBP$-$ & $4\hbar\omega$ &  1.73 & 0.07 \\
$2^{+}_{2}$ &  YSOX & $2\hbar\omega$ &  0.31 & -0.20\\
$2^{+}_{2}$ &  YSOX & $4\hbar\omega$ &  0.89 & -0.15\\
$2^{+}_{2}$ &  YSOX+& $4\hbar\omega$ &  0.26 & -0.21\\
$2^{+}_{2}$ &  WBP$-$ & $4\hbar\omega$ &  0.28 & -0.07\\
$2^{+}_{3}$ &  YSOX & $2\hbar\omega$ &  0.43 & -0.13\\
$2^{+}_{3}$ &  YSOX & $4\hbar\omega$ &  0.72 & -0.14\\
$2^{+}_{3}$ &  YSOX+& $4\hbar\omega$ &  0.46 & -0.19\\
$2^{+}_{3}$ &  WBP$-$ & $4\hbar\omega$ &  0.26 & -0.13\\
\bottomrule
\end{tabular*}
\end{center}

Such problem can be partially understood through the detailed investigation on the transition matrix elements and the configurations. Table~\ref{BGTdetail} presents the most important transition matrix element M$_{0p_{1/2} \rightarrow 0p_{3/2}}$ of B(GT)($^{14}$N$_{g.s.}$ $\rightarrow$ $^{14}$C$_{2^{+}_{1,2,3}}$) value. It is seen that all calculations give very small transition matrix elements between other orbits for these three B(GT) values. All shell-model results in Fig.~\ref{BGT1} give rather small B(GT)($^{14}$N$_{g.s.}$ $\rightarrow$ $^{14}$C$_{2^{+}_{2,3}}$) except those from YSOX in $4$ and $6\hbar\omega$ model spaces. The main difference comes from the M$_{0p_{1/2} \rightarrow 0p_{3/2}}$ term, which is much enhanced in the calculation from YSOX in $4\hbar\omega$ model space,
while the same transition matrix element in B(GT)($^{14}$N$_{g.s.}$ $\rightarrow$ $^{14}$C$_{2^{+}_{1}}$) is smaller in the same set of calculation.

%\begin{figure*}
%\includegraphics[scale=0.6]{occupancy.eps}
%\caption{\label{occupancy} (Color online) Comparison of the occupancies of the $0^{+}_{1,2}$ and $2^{+}_{1,2}$ states in $^{14}$C. The left four groups are from YSOX and YSOX$+$ results. The order in each group is YSOX in $2\hbar\omega$, YSOX$+$ in $2\hbar\omega$, YSOX in $4\hbar\omega$, and YSOX$+$ in $4\hbar\omega$, respectively.  The right four groups are from WBP$-$ and WBP results. The order in each group is WBP$-$ in $2\hbar\omega$, WBP in $2\hbar\omega$, WBP$-$ in $4\hbar\omega$, and WBP in $4\hbar\omega$, respectively.}
%\end{figure*}

Table.~\ref{configuration} shows the percentages of the $0$, $2$, and $4\hbar\omega$ configurations and the occupancies of each orbit in the $0^{+}_{1,2}$ and $2^{+}_{1,2,3}$ states of $^{14}$C. Calculations from YSOX in $4\hbar\omega$ model space give stronger mixing between $0$ and $2\hbar\omega$ configurations in all three $2^{+}$ states than those from all other calculations. A larger (smaller) $0\hbar\omega$ configuration in $2^{+}_{2,3}$ ($2^{+}_{1}$) states lead to a larger (smaller) M$_{0p_{1/2} \rightarrow 0p_{3/2}}$ terms in Table~\ref{BGTdetail}. From YSOX+ to YSOX, the reduction in the $\langle pp \mid$V$\mid sdsd\rangle$ interaction makes $0$ and $2\hbar\omega$ configurations more attractive in some states, such as $2^{+}_{1}$ state, but more repulsive in other states, such as $0^{+}_{1}$ state. When the interaction is reduced, the repulsive and attractive terms shown in Fig.\ref{ppsdsd} contribute to the attraction and repulsion of $0$ and $2\hbar\omega$ configurations, respectively. The spin dependent nature of the nuclear interaction differs $0^{+}_{1}$ and $2^{+}_{1}$ states.

The $sd$ shell neutrons in $2^{+}_{1}$ state mainly occupy $0d_{5/2}$ orbit. Some important TBME, $\langle p_{1/2}p_{1/2} \mid$V$\mid d_{5/2}d_{5/2}\rangle$ and $\langle p_{3/2}p_{3/2} \mid$V$\mid d_{5/2}d_{5/2}\rangle$, are repulsive and contribute to the enhanced occupancy on $0d_{5/2}$ orbit from YSOX+ to YSOX. The strong mixing between $0$ and $2\hbar\omega$ configurations in $2^{+}_{1,2}$ is also suggested based on the analysis of inelastic pion scattering~\cite{Fortune2016}. It should be noted that the phenomenological shell-model approaches normally gives less multi-$\hbar\omega$ mixing compared with the NCSM based methods. Because a phenomenological Hamiltonian is normally fitted with the assumption that $0\hbar\omega$ states are dominant states in most nuclei considered in the model space. For example, NCSM gives $56\%$ and $51\%$ of $0\hbar\omega$ configurations for the ground states of $^{12}$C and $^{16}$O, respectively \cite{Dytrych2007} and NCSpM gives around $65\%$ of $0\hbar\omega$ configurations for the former state \cite{Dreyfuss2013}, while YSOX in $4\hbar\omega$ model space gives $86\%$ and $69\%$ of $0\hbar\omega$ configurations for these two states, respectively.

\vspace{0mm}
\end{multicols}
\begin{center}
\tabcaption{\label{configuration}The configurations and the neutron occupancies of the $0^{+}_{1,2}$ and $2^{+}_{1,2,3}$ states in $^{14}$C.}
\begin{tabular*}{170mm}{@{\extracolsep{\fill}}ccccccccccc}
\toprule  State & Hamiltonian & Space & $0\hbar\omega$ (\%) & $2\hbar\omega$ (\%) & $4\hbar\omega$ (\%)& N$_{p_{1/2}}$& N$_{p_{3/2}}$& N$_{d_{3/2}}$& N$_{d_{5/2}}$& N$_{s_{1/2}}$\\
\hline
$0^{+}_{1}$ &  YSOX & $2\hbar\omega$ &  83.39& 16.61 & -    & 1.87 &	3.91 &	0.07 &	0.13 &	0.02\\
$0^{+}_{1}$ &  YSOX & $4\hbar\omega$ &  77.95& 20.90 & 1.15 & 1.81 &	3.87 &	0.09 &	0.20 &	0.04\\
$0^{+}_{1}$ &  YSOX+& $4\hbar\omega$ &  69.30& 28.15 & 2.55 & 1.75 &	3.81 &	0.12 &	0.28 &	0.04\\
$0^{+}_{1}$ &  WBP$-$ & $4\hbar\omega$ &  89.69& 10.04 & 0.27 & 1.91 &	3.96 &	0.03 &	0.08 &	0.02\\
$0^{+}_{2}$ &  YSOX & $2\hbar\omega$ &  3.60 & 96.40 & -    & 0.69 &	3.40 &	0.17 &	1.08 &	0.66\\
$0^{+}_{2}$ &  YSOX & $4\hbar\omega$ &  6.01 & 86.92 & 7.07 & 0.75 &	3.33 &	0.18 &	0.97 &	0.78\\
$0^{+}_{2}$ &  YSOX+& $4\hbar\omega$ &  6.12 & 82.28 & 11.60& 0.77 &	3.29 &	0.16 &	0.78 &	1.01\\
$0^{+}_{2}$ &  WBP$-$ & $4\hbar\omega$ &  2.01 & 94.78 & 3.21 & 0.71 &	3.33 &	0.16 &	1.25 &	0.56\\
$2^{+}_{1}$ &  YSOX & $2\hbar\omega$ &  75.13& 24.87 & -    & 1.78 &	3.87 &	0.09 &	0.23 &	0.04\\
$2^{+}_{1}$ &  YSOX & $4\hbar\omega$ &  49.12& 47.47 & 3.41 & 1.42 &	3.67 &	0.14 &	0.63 &	0.15\\
$2^{+}_{1}$ &  YSOX+& $4\hbar\omega$ &  60.22& 36.07 & 3.71 & 1.66 &	3.75 &	0.15 &	0.38 &	0.06\\
$2^{+}_{1}$ &  WBP$-$ & $4\hbar\omega$ &  81.30& 18.03 & 0.67 & 1.84 &	3.91 &	0.06 &	0.17 &	0.03\\
$2^{+}_{2}$ &  YSOX & $2\hbar\omega$ &  1.86 & 98.14 & -    & 0.66 &	3.39 &	0.18 &	1.10 &	0.67\\
$2^{+}_{2}$ &  YSOX & $4\hbar\omega$ &  15.11& 78.36 & 6.53 & 0.90 &	3.41 &	0.18 &	0.90 &	0.63\\
$2^{+}_{2}$ &  YSOX+& $4\hbar\omega$ &  1.26 & 86.72 & 12.02& 0.70 &	3.26 &	0.20 &	1.13 &	0.71\\
$2^{+}_{2}$ &  WBP$-$ & $4\hbar\omega$ &  2.21 & 94.09 & 3.70 & 0.71 &	3.33 &	0.15 &	1.25 &	0.56\\
$2^{+}_{3}$ &  YSOX & $2\hbar\omega$ &  3.92 & 96.08 & -    & 0.81 &	3.29 &	0.12 &	1.50 &	0.28\\
$2^{+}_{3}$ &  YSOX & $4\hbar\omega$ &  10.41& 83.11 & 6.48 & 0.92 &	3.27 &	0.13 &	1.45 &	0.23\\
$2^{+}_{3}$ &  YSOX+& $4\hbar\omega$ &  4.52 & 84.73 & 10.75& 0.83 &	3.21 &	0.14 &	1.58 &	0.24\\
$2^{+}_{3}$ &  WBP$-$ & $4\hbar\omega$ &  2.30 & 93.81 & 3.89 & 0.82 &	3.21 &	0.13 &	1.46 &	0.38\\
\bottomrule
\end{tabular*}
\end{center}
\begin{multicols}{2}

In light nuclei, there may be $\alpha$-cluster structure occurring. One of the famous example is the Hoyle state. It is shown that the Hoyle state demands $4-14\hbar\omega$ states in a NCSpM study \cite{Dreyfuss2013}, which is difficult to be described in the present phenomenological approach (effective Hamiltonians normally can not give the proper position of the Hoyle state). Because of the two more neutrons, low-lying states of $^{14}$C should be more dominated by the nucleon(s) excitation, rather than the $3\alpha$ structure of the Hoyle state. Thus the YSOX $0-4\hbar\omega$ results are not much different from that with the $0-6\hbar\omega$ model space. Although the $3\alpha$ structure may be excluded in the low-lying states of $^{14}$C, it is still difficult to identify whether there are $\alpha$ structure in these states in the present approach.

The effective single-particle energies~\cite{otsuka20012} in these states can be considered with the occupancies obtained from the shell model, as they are shown for the neutron rich C, N, and O isotopes~\cite{yuan20122}. The differences on the occupancies presented in Table.~\ref{configuration} among different sets of the calculations thus change the single-particle structures. The single-particle structure and the occupancy are normally dominated by the diagonal TBME, especially the monopole terms. The effect of the off-diagonal TBME on the occupancy is shown in the present work as a special example. It should be emphasized that the off-diagonal cross-shell interaction may have more effects on the configuration mixing beyond the configurations and occupancies, which demands further investigations. The uncertainty of a theoretical model normally consists two parts, statistical uncertainty from the not well determined parameters and systematic uncertainty from the deficiencies of the model. Recently, the two uncertainties of the liquid drop model are analysed based on the uncertainty decomposition method~\cite{yuan2016}. The present work provides a preliminary analysis on the systematic uncertainty of the shell model, which comes from the constrain on the model space.

\section{Summary}

In summary, the present work investigates the effect of the off-diagonal cross-shell interaction on the level and the transition rates of $^{14}$C. Based on the two well defined Hamiltonians in the $psd$ shell, WBP and YSOX, the observed excitation energies of the first few positive states in $^{14}$C can be well described in the $0-4\hbar\omega$ model space. A weaker strength of the $\langle pp \mid$V$\mid sdsd\rangle$ interaction is needed for WBP, while no changes are necessary for YSOX. It should be mentioned that the strength of the $\langle pp \mid$V$\mid sdsd\rangle$ interaction is not considered in the construction of WBP.

The B(GT) transition rates between the ground state of $^{14}$N and the positive states of $^{14}$C can be generally described by YSOX in the $0-4\hbar\omega$ or a larger model space under the framework of the nuclear shell model, but other Hamiltonians fail. Although it is hard to fully understand such results, further comparisons on the transition matrix elements and the configurations show that the partial reason comes from the strong mixing between the $0$ and $2\hbar\omega$ configurations in the $2^{+}$ states of $^{14}$C from YSOX results. The strong mixing is contributed by both the increment of the model space and the reduction of the off-diagonal cross-shell interaction. More effects may exist because of the complexity of the way that the off-diagonal TBME drives the nuclear structure.

The effect of the off-diagonal cross-shell interaction on the nuclear structure is not well investigated because it is less ``visible''. The present work shows some examples of its effect on the structure of $^{14}$C, while the Hamiltonians are constructed from the globally nice description on the nearby nuclei. It is of great importance to perform more investigations on this interaction based on both the phenomenological and realistic approaches.

\end{multicols}

\vspace{15mm}

\begin{multicols}{2}

\end{multicols}

\vspace{-1mm}
\centerline{\rule{80mm}{0.1pt}}
\vspace{2mm}

\begin{multicols}{2}

\end{multicols}

\clearpage
\end{CJK*}
\end{document}